\documentclass[10pt]{article}
 
\usepackage{fullpage}
\usepackage[all=normal,bibliography=tight]{savetrees}
\usepackage{microtype}
\usepackage{todonotes}
\usepackage{amsthm}
\usepackage{amssymb,bbm}
\usepackage{microtype}
\usepackage{multirow}

\usepackage{booktabs}
\usepackage{tabularx}
\usepackage{dcolumn}
\usepackage{colortbl}

\newcolumntype{d}[1]{D{.}{.}{#1}} 
\newcolumntype{A}{>{\columncolor[gray]{0.9}}d{1.3}}

\usepackage{colortbl}

\usepackage{xspace}
\usepackage{todonotes}
\usepackage{comment}
\usepackage[shortlabels]{enumitem}
\usepackage{tikz}
\usepackage{graphicx}
\usetikzlibrary{arrows, decorations.markings}
\usetikzlibrary{fit}					
\usetikzlibrary{backgrounds}

\usepackage{amsmath,amssymb,amsthm}
\usepackage{graphicx}
\usepackage{caption}
\usepackage{subcaption}
\usepackage{url}
\usepackage{mathtools}

\newcommand{\uselipics}{no}

\numberwithin{equation}{section}

\theoremstyle{definition}

\usepackage{multirow}
\usepackage{mathtools}

\newcommand{\iflipics}[2]{\ifthenelse{\equal{\uselipics}{yes}}{#1}{#2}}

\def\cqedsymbol{\ifmmode$\lrcorner$\else{\unskip\nobreak\hfil
\penalty50\hskip1em\null\nobreak\hfil$\lrcorner$
\parfillskip=0pt\finalhyphendemerits=0\endgraf}\fi} 

\newcommand{\mc}{\mathcal}
\newcommand{\Oh}{\mc{O}}

\def\grad_#1{\nabla\!_{#1}}
\def\topgrad_#1{\widetilde \nabla\!_#1}

\makeatletter
\newcommand{\manuallabel}[2]{\def\@currentlabel{#2}\label{#1}}
\makeatother

\title{Experimental evaluation of kernelization algorithms
to \textsc{Dominating Set}}

\author{Wojciech Nadara\thanks{%
Institute of Informatics, University of Warsaw, Poland.
\texttt{w.nadara@mimuw.edu.pl}
Supported by the ``Recent trends in kernelization: theory and experimental evaluation'' project, carried out within the Homing programme of the Foundation for Polish Science co-financed by the European Union under the European Regional Development Fund.}}

\date{}
 
\begin{document}

\maketitle

\begin{abstract}
The theoretical notions of graph classes with \textit{bounded expansion} and that are \textit{nowhere dense}
are meant to capture structural sparsity of real world networks that can be used to design 
efficient algorithms.

In the area of sparse graphs, the flagship problems are \textsc{Dominating Set} and its generalization $r$-\textsc{Dominating Set}.
They have been precursors for model checking of first order logic on sparse graph classes.
On class of graphs of bounded expansions the $r$-\textsc{Dominating Set} problem admits
a constant factor approximation, a fixed-parameter algorithm, and an efficient preprocessing routine: the so-called
linear kernel. This should be put in constrast with general graphs where \textsc{Dominating Set} is APX-hard
and $W[2]$-complete. 

In this paper we provide an experimental evaluation of kernelization algorithm
for \textsc{Dominating Set} in sparse graph classes
and compare it with previous approaches designed to the preprocessing for \textsc{Dominating Set}.
\end{abstract}

\section{Kernelization of \textsc{Dominating Set}} \label{sec:domset-theory}

In the $r$-\textsc{Dominating Set} problem, 
we are given a graph $G$ and a parameter $r$. Our goal is to find
smallest size of a set of vertices $D$ such that
for every vertex $v \in V(G)$ there exists a vertex $d \in D$ such that
the distance between $v$ and $d$ in $G$ is at most $r$.
Sets $D$ fulfilling this condition (not necessarily
with the smallest size) will be called dominating sets.
\textsc{Dominating Set} problem is a special case of this problem
where $r=1$. In decision versions of this problem we take additional parameter
$k$ in input and are asked whether size of smallest possible dominating set 
is at most $k$.

In arbitrary graphs the \textsc{Dominating Set} problem on $n$-vertex graphs is unlikely to be approximable in polynomial time within factor $(1-\varepsilon)\ln n$ for any $\varepsilon > 0$~\cite{Feige98},
$W[2]$-hard when parameterized by the solution size~\cite{DowneyF95} (and thus unlikely to be fixed-parameter tractable), and does not admit an algorithm running in time $\Oh(n^{k-\varepsilon})$ for any $\varepsilon > 0$
unless the Strong Exponential Time Hypothesis fails~\cite{PatrascuW10}. 
This is in stark contrast with planar graphs where \textsc{Dominating Set} admits a PTAS via the classic Baker's technique, is fixed-parameter tractable purely due to the degeneracy of planar graphs~\cite{AlonG09},
and --- what is most relevant for this work --- admits an efficient preprocessing routine: the so-called \emph{linear kernel}~\cite{AlberFN04}. 
That is, one can in polynomial time compress an input $(G,k)$ to the (decision version of the) \textsc{Dominating Set} problem on planar graphs to an equivalent instance $(G',k')$ with $k' \leq k$ and
$|G'| = \Oh(k')$. 

The preprocessing routine of~\cite{AlberFN04} consists of a number of relatively simple and local reduction rules, and thus is easy to implement. 
In~\cite{AlberFN04} and a follow-up work~\cite{AlberBN06} an experimental study of these rules is conducted on random planar graphs, showing that they can be preprocessed very efficiently for \textsc{Dominating Set}.

Since the work of~\cite{AlberFN04}, the study in subsequent years showed that the (parameterized) tractability of \textsc{Dominating Set} is not only limited to planar graphs, but applies to much wider sparse graph classes.
Here the crucial notions are that of \emph{bounded expansion}
and \emph{nowhere dense}, introduced by Ne\v{s}et\v{r}il and Ossona de Mendez~\cite{Sparsity}, that aim at capturing structural sparsity of real world networks.
Establishing fixed-parameter tractability of $r$-\textsc{Dominating Set} in these graph classes turned out to be a cornerstone result for a later proof that these graph classes are exactly the boundaries of tractability
of model checking of first order logic~\cite{GroheKS17}.

Recent results of Drange et al.~\cite{DrangeDFKLPPRVS16} and Eickmeyer et al.~\cite{EickmeyerGKKPRS17} show that bounded expansion and nowhere dense graph classes are also
limits for efficient preprocessing (so-called linear and polynomial kernels) for $r$-\textsc{Dominating Set}. 
Both these works treat a slight generalization of the problem where only a subset $Z \subseteq V(G)$ is required to be dominated and their
essence lies in introducing new reduction rule that is capable of identifying a vertex $v \in Z$ whose removal from $Z$ does not change the answer for the problem.

This reduction rule can be made oblivious to the actual graph class where it is applied (only the proof that it leads to a small kernel involves the assumption on the graph class).
Inspired by the experimental results of~\cite{AlberFN04,AlberBN06}, in this work we perform an experimental evaluation of this new reduction rule in random planar graphs
and in a number of real-worls sparse graphs using the benchmark from~\cite{NadaraPRRS18}.

Instead of treating the general $r$-\textsc{Dominating Set} problem, 
 we choose to evaluate algorithms to classical ($1$-)\textsc{Dominating Set} only, for the following reasons:
\begin{enumerate}
\item There has been previous experimental work on this topic~\cite{AlberFN04,AlberBN06}, so we can compare our results to already tested approaches. 
\item For $r=1$ we have a few easy reductions
which will help us a lot in reducing our graphs. 
\item The linear kernel for $r$-\textsc{Dominating Set} from~\cite{EickmeyerGKKPRS17} is a quite involved algorithm
where a few blowups of parameters appear. Because of that it seems that
this approach is rather impractical. However the earlier work of Drange et al.~\cite{DrangeDFKLPPRVS16} 
provides a different algorithm for linear kernel for $1$-\textsc{Dominating Set}
which is more ,,lightweight'' than the more general one.
\end{enumerate}

Throughout all of our experiments we will consider a bit generalized version
of \textsc{Dominating Set}, where each vertex can be one of two colors, either white or black.
Our goal is to pick smallest number of vertices so that all black vertices are dominated.
Black vertices serve role of both dominators and dominatees, whereas white vertices
serve role of dominators only. If we denote by $Z$ the set of black vertices
then by $dom(G, Z)$ we denote size of solution to such problem.
If all vertices are black then we get classical version of \textsc{Dominating Set}
what is expressed by equality $dom(G) = dom(G, V(G))$.

Formally speaking, since what we get at the end is an instance of problem
where some vertices are black and some are white, this is not precisely
an instance of typical \textsc{Dominating Set}.
However this can be easily fixed by adding two vertices
$u$ and $v$, where we put an edge $(u, v)$ and $(v, w)$ for every white vertex $w$,
declare all vertices in our graph as black and increase target size of dominating set by one.
Since $u$ has degree $1$, it is optimal to put $v$
into solution, so these two instances are equivalent and new one is an instance of typical \textsc{Dominating Set}. 

\section{The evaluated algorithm}
Let us try to express the gist of algorithm from~\cite{DrangeDFKLPPRVS16} in relatively short space.
Let us introduce a notion of \textit{strict core}. We say that $Z \subseteq V(G)$
is a strict core of $G$ if $dom(G) = dom(G, Z)$. It is obvious that $Z = V(G)$
is a strict core.

The main loop of the algorithm maintains a set $Z$ of black vertices that
is guaranteed to be a strict core.
In one step, it looks for a vertex $z \in Z$ such that $Z \setminus \{z\}$ is a strict core as well.
For such $z$ (which will be called
an \textit{irrelevant dominatee}), the algorithm recolors $z$ white and restarts.

By $N(u)$ we will denote set of $u$'s neighbors. By $N[u]$ we will denote
$N(u) \cup \{u\}$. We can also define $N[U]$ as $\bigcup_{u \in U} N[u]$.
We will say that set $B$ is $1$-scattered if
$$\underset{\substack{u \neq v,\\u, v \in B}}{\forall} N[u] \cap N[v] = \emptyset.$$ 

\paragraph{Reduction rule description.}
Let $Z$ be our current strict core (which are our black vertices).
Assume that we have a set $D$ and set of vertices $B \subseteq Z$ which is $1$-scattered in $G-D$ and
so that all vertices from $B$ have equal and nonempty set of neighbors in $D$, which we denote henceforth by $X$.
Let $M$ be $N[B] - D$ and $R$ be $N[M] \cap Z$. $M$ is the set of vertices outside of $M$ that can
possibly dominate $B$ and $R$ is the set of all black vertices that vertices from $M$ can dominate.
Assume that $dom(G, R) + 2 \le |B|$. Then any vertex of $B$ is an irrelevant dominatee.

\paragraph{Proof of rule's correctness.}
Let us now prove correctness of the rule above. Let $z$ be any vertex of $B$. We want to prove
that $dom(G, Z \setminus \{z\}) = dom(G, Z)$ (which is equal to $dom(G)$).
Let $E$ be any optimal dominating set of $Z \setminus \{z\}$. If it contains any vertex of $X$
then it dominates whole $B$ and in particular $z$, so it dominates $Z$, so this case is resolved.
So now we consider case when $E \cap X = \emptyset$. Since vertices of $B \setminus \{z\}$
are not dominated from $X$ then each of them needs to have its private dominator in $M$.
Let set of their dominators be called $C$. We know that $|C| \ge |B \setminus \{z\}| = |B| - 1 >
dom(G, R)$. Since $C \subseteq M$ we have that $N[C] \cap Z \subseteq N[M] \cap Z = R$. However we know
that we can dominate $R$ in a cheap way. If we delete $C$ from our solution and
put $dom(G, R)$ vertices that dominate $R$ we know that all vertices that were dominated
are still dominated (because $N[C] \cap Z \subseteq R$) and that size of our solution became smaller.
Therefore we arrive at a contradiction in that case with assumption that $E$
does not contain any vertex from $X$. This concludes proof of the correctness of this reduction rule.

\paragraph{Practical application of that rule.}
In original paper it is proved that if we are dealing with graphs
from class $\mathcal{G}$ with bounded expansion and $Z$ is sufficiently large (larger than
$ck$, where $k$ is number so that we are asked whether $dom(G) \le k$ and $c$ is
some constant depending on parameters of $\mathcal{G}$) then we can 
always apply that rule in polynomial time. However, we strived for making this approach as practical as possible,
so we propose a different way of looking for places where we can apply this rule.
\begin{enumerate}
\item We firstly greedily compute a dominating set of $Z$ which will be called $D'$.
\item Then we apply one of the uniform quasi-wideness algorithm on graph $G -D'$ and its subset 
$Z - D'$ to get a big $1$-scattered set $F$ after deleting small set $S$; we use the champion of
experimental evaluation of uniform quasi-wideness algorithms from~\cite{NadaraPRRS18} for that purpose.
\item We let $D = D' \cup S$, so that $F$ is $1$-scattered in $G - D$.
\item We group vertices of $F$ according to $(N(f) \cap D, N(N(f)) \cap D \cap Z)$, where $f \in F$.
\item For every such group $B$ of vertices with equal neighborhoods in $D$ and second neighborhood in $D \cap Z$
we greedily compute some dominating set of $R$, where $R = N[M]$ and $M = N[B] - D$.
\item If this greedy procedure returned set of size $g$ we can take any $max(0, |B| - g - 1)$
vertices of $B$ and recolor them to white.
\item After processing all these groups we look at vertex $v$ with greatest $|N(v) \cap D'|$.
If this value is at most $7$ then we end this procedure and otherwise 
we let $D' \coloneqq D' \cup \{v\}$ and we return to step 2.
\end{enumerate}
Note that we can use this reduction in parallel to many groups. We do not have to restart
whole procedure if one group yielded an irrelevant dominatee and if we see
that $|B| - g - 1$ is bigger then $1$ then we can whiten that many vertices at once, not only one of them.

\section{Other used reduction rules}
Apart from that reduction we used a set of simple ones and
two reduction rules introduced by Alber et al. in~\cite{AlberFN04,AlberBN06}.

Let us describe these simple rules:
\begin{enumerate}
\item If there is an edge between two white vertices, remove it.
\item If there is a white vertex $v$ so that it has no black neighbors, remove it.
\item If there is a white vertex $u$ and vertex $v$ such that $u$ and $v$ are adjacent and $N[u] \cap Y \subseteq N[v] \cap Y$
then remove $u$ (in fact previous rule is a special case of this one for non-isolated vertices.
\item If $u$ and $v$ are adjacent and $N[u] \supseteq N[v]$ and $v$ is black then $u$ can be colored white.
\item If there exists an isolated black vertex, put it into solution.
\item If there exists a black vertex with only one neighbor $u$, put $u$ into solution.
\end{enumerate}
Note that whenever we put any vertex into solution, we remove it from graph and make all its neighbors white.

It may seem not intuitive, but in fact these rules are more powerful than they seem.
For example, these rules alone are sufficient to fully solve (which means, reduce to empty graph)
significant part of random maximum planar graphs with $300$ vertices.

As already mentioned, we also use reduction rules proposed by Alber et al.,
we refer description of them to original papers~\cite{AlberFN04,AlberBN06}. 
Apart from a number of simple rules subsumed by the rules discussed above, Alber et al.~\cite{AlberFN04} introduce
two rules, one operating in a neighborhood of a single vertex, and one operating in a neighborhood of two vertices.
However one detail to note is that
since we operate on graphs with vertices that are black and white, not only black ones,
these rules need to be slightly adjusted, but doing this is straightforward. On the other hand,
when these rules deduce that it is optimal for some vertex to be picked as dominator
we can simply put it into solution, remove it and whiten its neighbors instead
of enforcing this artificially by putting gadget vertices as in original version.

\section{Tested approaches}
In all our evaluated algorithms we included all simple rules enumerated in previous section.
We chose six algorithms to evaluation which are identified by whether they use or not sparsity reduction
and by which subset of the two reduction rules of~\cite{AlberFN04} they use.
However we decided to drop approaches where we use reduction in the neighborhood of two vertices
but not the one in the neighborhood of the single vertex as this set seems unnatural. We call these approaches
\texttt{off.none, off.one, off.both, on.none, on.one, on.both}, where \texttt{on/off} stands for
whether sparsity reduction was used and \texttt{none/one/both} stands for what set
of Alber et al. reductions we used (\texttt{one} means just reduction for a neighborhood of a single vertex).

\section{Experimental setup, Hard- and Software}

The experiments on kernelization of \textsc{Dominating Set} has been performed on
an Asus K53SC laptop
with Intel® Core™ i3-2330M CPU @ 2.20GHz x 2 processor and with 7.7GiB of
RAM.

All implementation has been done in C++.
The code is available at~\cite{domset-repo}.

\section{Test data}

\paragraph{Random planar graphs.}
The first part of our benchmark dataset are random planar graphs generated by LEDA library
This allows us to compare our results with these presented in \cite{AlberFN04,AlberBN06}.
We try to distinguish planar graphs by their number of edges and by their edge density,
so we distinguish groups with $900$ and $6000$ edges and we distinguish groups with densities approximately $3, 2$ and $1.5$
(first group in our tests
are more precisely maximum planar graphs, where $|E(G)| = 3|V(G)| - 6$ holds).
As a result we get six groups with following sets of parameters $(|V(G)|, |E(G)|)$:
\begin{enumerate}
\item $(302, 900)$
\item $(450, 900)$
\item $(600, 900)$
\item $(2002, 6000)$
\item $(3000, 6000)$
\item $(4000, 6000)$
\end{enumerate}
In every group we generated $100$ graphs, so there are $600$ of them in total.

Note that LEDA can generate graphs with isolated vertices and they get lost in
conversion between formats (from LEDA format to list of edges), but any reasonable
algorithm deletes them at the beginning, so they don't matter much. However for
the sake of clarity we specify that we do not take into account such isolated vertices
when stating graph's size, so in fact real densities are a bit bigger than declared $1.5$ and $2$
in respective groups.

\paragraph{Real-world sparse graphs.}
We used also the benchmark set of real-world sparse graphs from different sources
gathered earlier for experimental evaluation of algorithms in graphs of bounded expansion
and nowhere dense graph classes~\cite{NadaraPRRS18}. 
We used two test groups from~\cite{NadaraPRRS18}: the ``small'' graphs, consisting
of graphs up to $1\,000$ edges, and the ``medium'' graphs, consisting of graphs
between $1\,000$ and $10\,000$ edges.
We refer to~\cite{NadaraPRRS18} for a detailed discusion of the composition of these benchmark sets.

\section{Results and discussion}

Full results for random planar graphs are presented in Table~\ref{tab:domset}
and for real-world sparse graphs are presented in Table~\ref{tab:domset2}.
Before proceeding to results description let us mention that experiments with these algorithms
focused on performance of different kinds of reduction rules
measured by whether they reduce our graph or not.
Rules were not implemented with focus on their running time and they were applied
exhaustively until no rule can be used. That is why we do not provide exact
running times of these algorithms.
However to give a rough estimate of order of magnitude, execution of all tested approaches on all tests took
more or less 1-2 hours in total.

\begin{table}[htb]
\begin{center}
\begin{tabular}{l|ll|l|l}
$|V|$ & sparsity & Alber et al. & remaining vtcs & remaining edges \\\hline
\multirow{6}{*}{302} & off & none & 0.0293709 & 0.0116333\\
 & on & none & 0.0293709 & 0.0116333\\
 & off & one & 0.0293709 & 0.0116333\\
 & on & one & 0.0293709 & 0.0116333\\
 & off & both & 0.00546358 & 0.0022\\
 & on & both & 0.00546358 & 0.0022\\
\hline
\multirow{6}{*}{450} & off & none & 0.0244906 & 0.0143444\\
 & on & none & 0.024468 & 0.0143222\\
 & off & one & 0.0237896 & 0.0138333\\
 & on & one & 0.0237896 & 0.0138333\\
 & off & both & 0.0064449 & 0.00402222\\
 & on & both & 0.0064449 & 0.00402222\\
    \hline
\multirow{6}{*}{600} & off & none & 0.0119285 & 0.0082\\
 & on & none & 0.0119285 & 0.0082\\
 & off & one & 0.0119285 & 0.0082\\
 & on & one & 0.0119285 & 0.0082\\
 & off & both & 0.00287069 & 0.00193333\\
 & on & both & 0.00299551 & 0.00201111\\
    \hline
\multirow{6}{*}{2002} & off & none & 0.0275624 & 0.010975\\
 & on & none & 0.0280769 & 0.011185\\
 & off & one & 0.0275624 & 0.010975\\
 & on & one & 0.0278971 & 0.0111083\\
 & off & both & 0.00655345 & 0.00272333\\
 & on & both & 0.00655345 & 0.00272333\\
     \hline
\multirow{6}{*}{3000} & off & none & 0.0221042 & 0.0128967\\
 & on & none & 0.0222806 & 0.0130083\\
 & off & one & 0.0214934 & 0.0124533\\
 & on & one & 0.0215884 & 0.0125067\\
 & off & both & 0.00456333 & 0.00274667\\
 & on & both & 0.00455315 & 0.00274167\\
     \hline
\multirow{6}{*}{4000} & off & none & 0.0116722 & 0.00798167\\
 & on & none & 0.0116722 & 0.00798167\\
 & off & one & 0.0115836 & 0.00789333\\
 & on & one & 0.0115836 & 0.00789333\\
 & off & both & 0.00253778 & 0.00174333\\
 & on & both & 0.00253778 & 0.00174333\\
\end{tabular}
\caption{Results of the experiments on dominating set kernelization in random planar graphs.
The last two columns indicate the average fraction of vertices and edges, respectively,
that remain in the reduced graph.}\label{tab:domset}
\end{center}
\end{table}

\begin{table}[htb]
\begin{center}
\begin{tabular}{l|ll|l|l}
$|V|$ & sparsity & Alber et al. & remaining vtcs & remaining edges \\
\hline \multirow{6}{*}{small}
 & off & none & 0.446102 & 0.413362 \\
 & on & none & 0.446102 & 0.413362 \\
 & off & one & 0.446102 & 0.413362 \\
 & on & one & 0.446102 & 0.413362 \\
 & off & both & 0.426819 & 0.398826 \\
 & on & both & 0.426819 & 0.398826 \\
\hline \multirow{6}{*}{medium}
 & off & none & 0.239025 & 0.218247 \\
 & on & none & 0.239025 & 0.218247 \\
 & off & one & 0.239025 & 0.218247 \\
 & on & one & 0.239025 & 0.218247 \\
 & off & both & 0.230227 & 0.211836 \\
 & on & both & 0.230227 & 0.211836 \\\hline
 \end{tabular}
\caption{Results of the experiments on dominating set kernelization in ``small'' and ``medium''
  datasets of real-world sparse graphs from~\cite{NadaraPRRS18}.
  Last two columns were computed as follows. For every test we take a fraction of remaining
 vertices and edges and then we take averages of these fractions.}\label{tab:domset2}
\end{center}
\end{table}

First, our results confirm the experimental findings of Alber et al.~\cite{AlberFN04,AlberBN06}
that their basic reduction rules are very powerful with respect to sparse graphs.
Furthermore, the main strength
of their reduction rules
lies in the combination of both rules (as opposed to the first rule alone):
  this effect is very strong in random planar graphs, but also visible in the results
  in real-world sparse graphs.

It turned out that in fact on vast majority of tests it does not make a difference
whether we apply sparsity reduction or not. However when looking at logs we
noted that in fact sparsity reduction is applied significantly more often than it may seem
than when looking at just results, however probably in many cases what sparsity reduction reduced
was easily discoverable by some combination of simple rules. Cases where sparsity reduced something
that was not covered by simple rules existed, however they were very rare. One surprising
observation when looking at these results is that using sparsity reduction sometimes
\textit{worsened} results. However there is a fine explanation behind why this could hypothetically be a case.
Note that in order for some rules to be applied we need to know that some vertex has some particular color.
And if sparsity reduction tells us that some vertex $v$ is black, but can be colored white
without changing an answer then it may look like a profit, but we shut a way for us to deduce something based on fact that
$v$ is black.

We now sketch how to mitigate this problem. 
Some white vertices are white because they were already dominated by some vertex that
was put into solution and some are white because some reduction rule told us that such vertex is an irrelevant dominatee.
We could introduce an intermediate gray color. It would mean that if $v$ is gray then there is no vertex from already
determined part of solution that dominates $v$, however we know that $v$ is an irrelevant dominatee, meaning
that any optimal solution to remaining instance also dominates $v$.
So in fact it would mean that in reduction rules we would be allowed to choose whether $v$ is black
or white. Evaluating such an enhanced approach is a topic for future work.

\bibliographystyle{plain}
\bibliography{refs}

\end{document}